\documentclass[
aip,
 amsmath,amssymb,
reprint, %
]{revtex4-1}
\usepackage{color}
\usepackage{graphicx}
\usepackage{dcolumn}
\usepackage{bm}

\begin{document}


\preprint{AIP/123-QED}

\title{Mixed random walks with a trap in scale-free networks including nearest-neighbor and next-nearest-neighbor jumps}

\author{Zhongzhi Zhang}
\email{zhangzz@fudan.edu.cn}
\homepage{http://www.researcherid.com/rid/G-5522-2011}

\author{Yuze Dong}

\author{Yibin Sheng}

\affiliation {School of Computer Science, Fudan University,
Shanghai 200433, China}

\affiliation {Shanghai Key Laboratory of Intelligent Information
Processing, Fudan University, Shanghai 200433, China}

\date{\today}

\begin{abstract}
Random walks including non-nearest-neighbor jumps appear in many real situations such as the diffusion of adatoms and have found numerous applications including PageRank search algorithm, however, related theoretical results are much less for this dynamical process. In this paper, we present a study of mixed random walks in a family of fractal scale-free networks, where both nearest-neighbor and next-nearest-neighbor jumps are included. We focus on trapping problem in the network family, which is a particular case of random walks with a perfect trap fixed at the central high-degree node. We derive analytical expressions for the average trapping time (ATT), a quantitative indicator measuring the efficiency of the trapping process, by using two different methods, the results of which are consistent with each other. Furthermore, we analytically determine all the eigenvalues and their multiplicities for the fundamental matrix characterizing the dynamical process. Our results show that although next-nearest-neighbor jumps have no effect on the leading sacling of the trapping efficiency, they can strongly affect the prefactor of ATT, providing insight into better understanding of random-walk process
in complex systems.
\end{abstract}




\date{\today}
\maketitle



\section{Introduction}

As a powerful tool for describing and studying complex systems, network science (complex networks) has attracted substantial attention of the scientific community in the past decade~\cite{AlBa02,Ne03}. A central problem in the field of complex networks is to understand the relationship between various structural properties and dynamical processes occurring on networks. Among many different dynamical processes, random walks are a fundamental natural process, since they describe or express a wealth of other physical processes, including navigation~\cite{Kl00}, search~\cite{GuDiVeCaAr02,BeLoMoVo11}, and so on. Thus far, random walks have found a plethora of applications in interdisciplinary fields~\cite{BadaViCa05, BrHuGe06, Le06, PoLa06,RoEsLaWeLa14, ChBa07}. In view of their theoretical and practical relevance, continuously increasing endeavors have been devoted to study random walks on complex networks~\cite{HaBe87,BuCa05,Vo11,BeVo14, BeGuVo15}.

One of the most important quantities related to random walks is first-passage time (FPT)~\cite{Re01,NoRi04}. The FPT from a source node $s$ to a target node $t$ is defined as the expected time for a walker starting at node $s$ to arrive at $t$ for the first time.
The mean of FPTs to a given target over all starting nodes is known as mean first-passage time (MFPT), which plays an essential role in various realistic situations, such as trapping problem~\cite{Mo69}, target search~\cite{JaBl01,BeLoMoVo11}, and lighting harvesting~\cite{BaKlKo97,BaKl98,BeHoKo03}. MFPT has been deeply studied in different networks~\cite{BeCoMo05,CoBeKl07,CoBeTeVoKl07,BiChKlMeVo10,HwLeKa12}, including the Sierpinski fractal~\cite{KoBa02,BeTuKo10,BaKo13}, the $T$-fractal~\cite{KaRe89,Ag08,LiWuZh10,WuZh13,PeXu14}, dendrimers~\cite{WuLiZhCh12,LiZh13JCP,PeZh14} and hyperbrached polymers~\cite{WuLiZhCh12,LiZh13JCP} square-planar lattices~\cite{GLKo05}, scale-free networks~\cite{ZhQiZhXiGu09,TeBeVo09,AgBu09,AgBuMa10,MeAgBeVo12}, as well as weighted networks~\cite{ZhShCh13,LiZh13PRE}.

Previous studies uncovered the critical effects of structure and weight of the underlying systems on MFPT, for example, inhomogeneous degree~\cite{ZhQiZhXiGu09,AgBu09} or weight~\cite{ZhShCh13,LiZh13PRE}. 
However, most existent works focus on nearest-neighbor random walks, neglecting the role of non-nearest-neighbor hopping, which has been implicated in some physical processes, such as exciton migration in crystals~\cite{SoBo72}, photosynthesis~\cite{Kn68}, and the surface diffusion of adatoms~\cite{EhSt80}. Particularly, a recent work pointed out the experimental evidence for and the physical significance of non-nearest-neighbor jumps in the diffusion of adatoms~\cite{AnEh07}. Due to its significant importance, non-nearest-neighbor hopping has been considered in various contexts~\cite{BeLoMoVo11,OsWiLiBu07,RoReBuWiOsLi10,BeLoMoVo08}. Nevertheless, in contrast to nearest-neighbor random walks, related research about MFPT for random walks including non-nearest-neighbor jumps is much less~\cite{BaKo13,MuKo83}. Even if the inclusion of non-nearest-neighbor jumps may not affect the scaling exponent of MFPT, we may expect that it can significantly modify the prefactor of MFPT~\cite{CoBeKl07,BiChKlMeVo10}. However, it is still not well understood how the prefactor changes with non-nearest-neighbor jumps.

In this paper, we study random walks in a family scale-free fractal networks~\cite{SoHaMa06,RoHaBe07} with a deep trap placed at the central large-degree node. During the process of random walks, both  nearest-neighbor and non-nearest-neighbor jumps are allowed with different probability controlled by a parameter. We obtain two expressions for the MFPT to the trap by using two different techniques, the results of which are consistent with each other. In addition, we find all the eigenvalues and their degeneracies of the fundamental matrix characterizing the trapping problem. 
The obtained result indicates that the prefactor of the MFPT to the target is dependent on the probability parameter, which shows that the inclusion of non-nearest-neighbor hopping has a vital influence on random walks in the networks under consideration.

\section{Network model\label{Sec:model}}

\begin{figure}
\begin{center}
\includegraphics[width=1.0\linewidth]{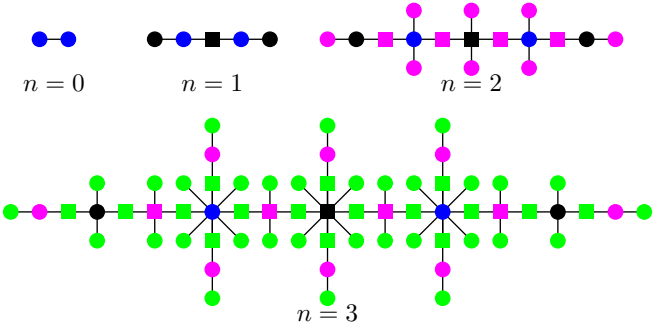} \
\end{center}
\caption[kurzform]{ The first two iterations of a
specific network for $m=1$.}
\label{network}
\end{figure}

The studied networks are defined in an iterative way. Let $F_n$
denote the networks after $n$ ($n\geq
0$) iterations. Then, $F_n$ are constructed as follows~\cite{SoHaMa06,RoHaBe07}. For $n=0$, $F_0$ contains two nodes linked by an edge. For $n\geq 1$, $F_n$ is
obtained from $F_{n-1}$ by performing the following
operations on every edge in $F_{n-1}$: replace the edge by a
path of 2 links long, with the two endpoints of the path being the
same endpoints of the original edge (the new node having an initial
degree 2 in the middle of path is referred to as an internal node),
then attach $m$ new nodes with an initial degree 1 (called external
nodes) to each endpoint of the path. Figure~\ref{network}
illustrates the construction process for a limiting case of $m=1$,
showing the first two iterative processes.

By construction, at each generation $n_i$ ($n_i\geq 1$) the number
of newly introduced nodes is $V_{n_i}=(2m+1)(2m+2)^{n_i-1}$, among which $(2m+2)^{n_i-1}$ nodes are internal nodes and the remaining $2m(2m+2)^{n_i-1}$ nodes are external
nodes. Then,
the total number of  nodes $N_n$ in $F_n$ is
$N_n=\sum_{n_i=0}^{n}V_{n_i}=(2m+2)^{n}+1$, and the total number of
edges is $E_n=N_n-1=(2m+2)^{n}$. Let $d_i(n)$ represent the degree of a node
$i$ in $F_n$, which was generated at iteration $n_i$ ($n_i\geq 0$). Then $d_i(n)=2(m+1)^{n-n_{i}}$ if $i$ is an internal node; and $d_i(n)=(m+1)^{n-n_{i}}$ if node $i$
is an external node. Hence, after each new iteration the degree of every node increases by $m$ times, i.e., $d_i(n)=(m+1)\,d_i(n-1)$.

This networks under consideration display the remarkable topological
features as observed in various real systems. They are scale free with their degree distribution $P(k)$ following a power law form $P(k)\sim k^{-\gamma}$,  where $\gamma=1+\ln (2m+2)/ \ln (m+1)$~\cite{RoHaBe07}. In addition, they are
fractal with the fractal dimension being $d_B=\ln (2m+2)/ \ln
2$~\cite{SoHaMa06}.

\section{Definition of mixed random walks \label{Sec:walk}}

We define a novel type of random walks taking place in the fractal scale-free networks $F_n$, which include both nearest-neighbor and next-nearest-neighbor jumps and is thus called mixed random walks hereafter. Let $A_n$ denote the adjacency matrix of $F_{n}$, which encodes the structure information of $F_{n}$. The entries $A_n(i,j)$ of $A_n$ are defined by $A_n(i,j)=1$ if nodes $i$ and $j$ are adjacent in $F_{n}$, or $A_n(i,j)=0$ otherwise. Note that all random walks are determined by their corresponding transition probability matrices. We use $P_n$ to represent the transition probability matrix for mixed random walks in $F_n$, whose entry $P_n(i,j)$ is the jump possibility from node $i$ to node $j$.

During the process of mixed random walks in $F_n$, if the current location of the walker is an old node, which is already existent in $F_{n-1}$, it is allowed to jump to both nearest neighbors and next nearest neighbors, with their respective probabilities are $\theta$ and $1-\theta$ ($0\leq \theta \leq1$); if the current state of the walker is at a new node created at iteration $n$, then it can only jump to nearest neighbors. In other words, for mixed random walks in $F_n$, the walker performs isotropic nearest-neighbor random walks in either $F_{n-1}$ or $F_n$, with respective probabilities $\theta$ and $1-\theta$. Concretely, for mixed random walks in $F_n$, the transition probability is defined by
\begin{equation}\label{T1}
P_{n} (i,j)=
\begin{cases} \frac{\theta A_{n} (i,j)}{d_i(n)}, & i\in\alpha, \, j\in\beta,\\
  \frac{(1-\theta) A_{n-1} (i,j)}{d_i (n-1)}, & i\in\alpha, \,j\in\alpha,\\
  \frac{A_{n} (i,j)}{d_i (n)}, & i\in\beta, \,j\in\alpha,\\
  \frac{A_{n} (i,j)}{d_i (n)},\ \ & i\in\beta, \,j\in\beta,
\end{cases}
\end{equation}
where $\alpha$ represents the set of nodes belonging to $F_{n-1}$, and $\beta$ represents the set of nodes generated at $n$th iteration.
Since all new nodes in $F_{n}$ are not adjacent, there is no transition between any pair of new nodes in $F_{n}$. Thus, Eq.~(\ref{T1}) is reduced to
\begin{equation}\label{T2}
P_{n} (i,j)=
\begin{cases} \frac{\theta A_{n} (i,j)}{d_i(n)}, & i\in\alpha, \, j\in\beta,\\
  \frac{(1-\theta) A_{n-1} (i,j)}{d_i (n-1)}, & i\in\alpha, \,j\in\alpha,\\
  \frac{A_{n} (i,j)}{d_i (n)}, & i\in\beta, \,j\in\alpha,\\
  0,\ \ & i\in\beta, \,j\in\beta.
\end{cases}
\end{equation}

There are two special cases for the above-defined mixed random walks in $F_n$.  For $\theta=0$, it reduces to random walks in $F_n$; for $\theta=1$, it is exactly random walks in $F_{n-1}$. As expected, the probability parameter $\theta$ dominates the process of mixed random walks in $F_n$. Below we will study a particular case of mixed random walks in $F_n$ with a deep trap positioned at the central hub node, i.e., the internal node generated at iteration 1, and show that the parameter $\theta$ significantly influences the ATT to the trap, as well as the eigenvalues of the fundamental matrix associated with the trapping problem.

\section{Mixed random walks with a perfect trap at a hub node}

In the sequel, we examine mixed random walks in $F_n$ with a perfect trap at the internal node created at the first generation. We will derive explicit formulas for the ATT to the target. Moreover, we will obtain the full spectra for the fundamental matrix describing the trapping problem. Based on these results, we will show that next-nearest-neighbor jumps, dominating by the parameter $\theta$, have a substantial effect on the dynamic process, especially the prefactor of ATT.

\subsection{Formulation of trapping problem}

The trapping problem is a kind of random walks with a deep trap fixed at a certain location. We here address mixed random walks in $F_n$  in the presence of a trap placed at the central node, that is, the unique internal node created at the first generation. For the convenience of description, we label all the nodes in $F_n$ as follows. For $n=1$, the trap node is labeled as 1, the initial two nodes belonging to $F_0$ are labeled as 2 and 3, and all other nodes are labeled as $4$, $5$, $\ldots$, $2m+3$. For each new generation $n>1$, only those new nodes generated at this generation are labeled, while the labels of all old nodes remain unchanged, i.e., we consecutively label new nodes in $F_n$ as $N_{n-1}+1$, $N_{n-1}+2$, $\cdots$, $N_n$.

Let $T_i^{(n)}$ represent the trapping time (TT) for a walker initially placed at node $i$ (other than the trap) to arrive at the trap node for the first time. In fact, $T_i^{(n)}$ is the FPT from node $i$ to the trap. Then, the average trapping time (ATT) $\langle T \rangle_n$, which is the mean of $T_i^{(n)}$ over all non-trap initial nodes in network $F_n$, is given
by
\begin{equation}\label{C4}
 \langle T
\rangle_n=\frac{1}{N_n-1}\sum_{i=2}^{N_n} T_i^{(n)}\,.
\end{equation}
By definition, the quantity ATT $\langle T\rangle_n$ is the MFPT to the trap, which is very important since it is a quantitative indicator measuring the trapping efficiency, with small ATT corresponding to efficient trapping process. Below we will study the two quantities $T_i^{(n)}$ and $\langle T \rangle_n$.

For $T_i^{(n)}$, it satisfies the following relation
\begin{equation}\label{C1}
T_i^{(n)}=\sum_{j=2}^{N_n}P_n(i,j)\,T_j^{(n)}+1\,,
\end{equation}
which can be rewritten in matrix form as
\begin{equation}\label{C2}
T= \bar{P}_n\,T+ e\,,
\end{equation}
where $T=\left(T_2^{(n)}, T_3^{(n)}, \cdots, T_{N_n}^{(n)}\right)^{\top}$ is an ($N_n-1$)-dimensional vector, $\bar{P}_n$ is a matrix of order $N_n-1$ that is actually a submatrix of $P_n$ with the row and column corresponding the trap being removed, and $e=(1, 1, \cdots, 1)^{\top}$ is the ($N_n-1$)-dimensional vector of all ones. Equation~(\ref{C2}) implies
\begin{equation}\label{C3}
T=(I-\bar{P}_n)^{-1} e=M_n \,e\,,
\end{equation}
where $I$ is the identity matrix of order $(N_n-1)\times (N_n-1)$. Matrix $M_n=(I-\bar{P}_n)^{-1}$  is often called  fundamental matrix~\cite{KeSn76} of the trapping problem. Equation~(\ref{C3}) means
\begin{equation}\label{C4}
T_i^{(n)}=\sum_{j=2}^{N_n}M_n(i,j)\,,
\end{equation}
where $M_n(i,j)$ denotes the $ij$th element of the fundamental matrix, which is the mean number of visits of node $j$ by the walker starting from node $i$ before being trapped.
Inserting Eq.~(\ref{C3}) into Eq.~(\ref{C4}) leads to
\begin{equation}\label{C5}
\langle T\rangle_n=\frac{1}{N_n-1}\sum_{i=2}^{N_n}\sum_{j=2}^{N_n}M_n(i,j)\,.
\end{equation}

Equation~(\ref{C5}) shows that the problem of evaluating ATT $\langle T \rangle_n$ can be reduced to computing the sum of all elements of the associated fundamental matrix. However, before finding the sum, one must first invert a matrix, which demands a large computational effort when the networks are very large. Thus, Eq.~(\ref{C5}) is only valid for those networks with a small number of nodes, but it generates exact results that can be used to check the results for ATT derived by other approaches. In what follows, we will analytically determine the closed-form expression for ATT $\langle T \rangle_n$ using another technique. Moreover, we will determine all the eigenvalues of the fundamental matrix $M_n$, the largest eigenvalue of which is proportional to the leading scaling of the ATT.

\subsection{Exact solution to average trapping time}

The particular selection of trap location and the special network structure allow to determine exactly the ATT $\langle T \rangle_n$ for arbitrary $n$. In order to evaluate $\langle T \rangle_n$, we use $\Lambda_n$ to represent the set of all nodes in $F_n$, and
use $\bar{\Lambda}_n$ to denote the set of those nodes created at generation $n$.
Thus, $\Lambda_n=\bar{\Lambda}_n \cup \Lambda_{n-1}$. For the convenience of
computation for $\langle T \rangle_n$, we introduce the following quantities for any $g \leq n$:
$T_{g,\text{tot}}^{(n)}=\sum_{i\in \Lambda_g} T_i^{(n)}$ and
$\bar{T}_{g,\text{tot}}^{(n)}=\sum_{i\in \bar{\Lambda}_g}
T_i^{(n)}$. Then,
\begin{equation}\label{ES1}
\langle T\rangle_n=\frac{1}{N_n-1}T_{g,\text{tot}}^{(n)}\,.
\end{equation}
The specific case $\theta=1$ has been studied in~\cite{ZhLiMa11}. For this case, we represent the quantities $T_i^{(n)}$, $T_{g,\text{tot}}^{(n)}$,
$\bar{T}_{g,\text{tot}}^{(n)}$, and $\langle T \rangle_n$  by $H_i^{(n)}$, $H_{g,\text{tot}}^{(n)}$,
$\bar{H}_{g,\text{tot}}^{(n)}$, and $\langle H \rangle_n$, respectively.
It has been proved~\cite{ZhLiMa11} that
\begin{eqnarray}\label{ES2}
 \langle H
\rangle_n
&=&\frac{4m^{2}+4m+1}{2(4m^{2}+7m+3)}2^{n}(2m+2)^{n}+\nonumber\\
&\quad&\frac{16m^{2}+16m+3}{4(4m^{2}+7m+3)}2^{n}-\frac{4m^{2}+4m+1}{2(4m^{2}+7m+3)}\,,
\end{eqnarray}
which is helpful for the following derivation.

Next, we show that there exists a useful relation between $T_i^{(n)}$ and $H_i^{(n-1)}$.
Let's examine a node $i$ in $F_n$. Notice that after one iteration, the degree of an old node $i$ in $F_{n-1}$ increases from $d_i(n-1)$ to
$(m+1)\,d_i(n-1)$. Moreover, all these $(m+1)\,d_i(n-1)$ neighbors of node $i$ are
new nodes created at iteration $n$, among which $m \,d_i(n-1)$
neighbors are external nodes, and the remaining $d_i(n-1)$ neighbors are
internal nodes. For mixed random
walks in $F_{n}$, let $X$ be the FPT for a
particle starting from node $i$ to any of its $d_i(n-1)$ old
neighbors, namely, those nodes directly connected to $i$ at
iteration $n-1$, and let $Y$ (resp. $Z$) be the FPT
for going from any of the $d_i(n-1)$ (resp. $m d_i(n-1)$) internal (resp. external) new neighbors of $i$ to one of its $d_i(n)$ old neighbors.
Then, $X$, $Y$, and $Z$ follow the relations:
\begin{eqnarray}\label{ES3}
\left\{
\begin{array}{ccc}
X&=&\frac{m \theta}{m+1}(1+Z) + \frac{\theta}{m+1}(1+Y)+(1-\theta)\,,\\
Y&=&\frac{1}{2} + \frac{1}{2}(1+X)\,,\\
Z&=&1+X\,.
 \end{array}
 \right.
\end{eqnarray}
Eliminating $Y$ and $Z$ in Eq.~(\ref{ES3}) yields $X=\frac{(2m+2)(1+\theta)}{(2-2\theta)m+(2-\theta)}$.
Therefore, when the networks evolve from iteration $n-1$ to iteration
$n$, the FPT from any node $i$ ($i \in F_{n-1}$) to another node
$j$ ($j \in F_{n-1}$) increases by a factor of $\frac{(2m+2)(1+\theta)}{(2-2\theta)m+(2-\theta)}$. Thus, we
have
\begin{equation}\label{ES4}
T_i^{(n+1)}=\frac{(2m+2)(1+\theta)}{(2-2\theta)m+(2-\theta)}\,H_i^{(n)},
\end{equation}
an expression useful for the following derivation of the exact solution to ATT $\langle T \rangle_n$.
For $\theta=1$, Eq.~(\ref{ES4}) becomes
\begin{equation}\label{ES5}
H_i^{(n+1)}=(4m+4)H_i^{(n)}\,.
\end{equation}

Equation~(\ref{ES1}) shows that, in order to determine $\langle T \rangle_n$, we can alternatively estimate  $T_{n, {\rm tot}}^{(n)}$ that obeys the following relation
\begin{eqnarray}\label{ES6}
T_{n, {\rm tot}}^{(n)} &=& T_{n - 1, {\rm tot}}^{(n)} + \bar{T}_{n,{\rm tot}}^{(n)}\nonumber\\
&=& \frac{(2m+2)(1+\theta)}{(2-2\theta)m+(2-\theta)}\,H_{n - 1, {\rm tot}}^{(n-1)} + \bar{T}_{n, {\rm tot}}^{(n)}.
\end{eqnarray}
Hence, to find $T_{n, {\rm tot}}^{(n)}$, it is necessary to first
explicitly determine the quantity $\bar{T}_{n, {\rm tot}}^{(n)}$.

For an arbitrary external node $i_{\rm ext}$ in $F_n$,
which was generated at iteration $n$ and linked to an old node $i$,
we have
\begin{equation}\label{ES7}
T_{i_{\rm ext}}^{(n)} = 1 + T_i^{(n)} \,.
\end{equation}
While for an arbitrary internal node $l^{ij}_{\rm int}$, which was created at generation $n$ and attached to a pair old nodes $i$ and $j$,
we have
\begin{equation}\label{ES8}
T_{l^{ij}_{\rm int}}^{(n)} = 1 + \frac{1}{2}T_i^{(n)} + \frac{1}{2}T_j^{(n)} \,.
\end{equation}
Then, by construction, it is easy to establish relation
\begin{eqnarray}\label{ES9}
& &\bar{T}_{n,{\rm tot}}^{(n)}\nonumber\\ &=&|\bar{\Lambda}_n| + \sum_{i \in \Lambda_{n-1}}\left[\left(m+\frac{1}{2}\right) d_i(n-1)\times T_i^{(n)}\right]\nonumber\\
&=& |\bar{\Lambda}_n| + \left(m+\frac{1}{2}\right)\times\nonumber\\
&\quad&\sum_{i \in \Lambda_{n-1}}\left( d_i(n-1)\frac{(2m+2)(1+\theta)}{(2-2\theta)m+(2-\theta)}H_i^{(n-1)}\right),
\end{eqnarray}
where $|\bar{\Lambda}_n|$ denotes the cardinality of set $\bar{\Lambda}_n$.
For $\theta=1$, Equation~(\ref{ES9}) reduces to
\begin{equation}\label{ES10}
\bar{H}^{(n)}_{n, {\rm tot}} =|\bar{\Lambda}_n| +\left(m+\frac{1}{2}\right) \sum_{i \in \Lambda_{n-1}}\left(d_i(n-1)(4m+4)H_i^{(n-1)}\right).
\end{equation}

Combining Eqs.~(\ref{ES9}) and~(\ref{ES10}), we obtain
\begin{equation}\label{ES11}
\frac{\bar{T}_{n, {\rm tot}}^{(n)}-|\bar{\Lambda}_n|}{\frac{(2m+2)(1+\theta)}{(2-2\theta)m+(2-\theta)}}=\frac{\bar{H}_{n, {\rm tot}}^{(n)}-|\bar{\Lambda}_n|}{4m+4},
\end{equation}
from which we can further derive
\begin{eqnarray}\label{ES12}
\bar{T}_{n, {\rm tot}}^{(n)}&=&\frac{(2m+2)^{n-1}(1+2m)(3+4m)(1-\theta)}{4+4m-2\theta-4m\theta}\nonumber\\
&\quad& +\frac{1+\theta}{4+4m-2\theta-4m\theta}\bar{H}_{n, {\rm tot}}^{(n)},
\end{eqnarray}
where $|\bar{\Lambda}_n|=(2m+1)(2m+2)^{n-1}$ was used.
On the other hand,
\begin{equation}\label{ES13}
\bar{H}_{n, {\rm tot}}^{(n)}={H}_{n, {\rm tot}}^{(n)}-{H}_{n-1, {\rm tot}}^{(n)}={H}_{n, {\rm tot}}^{(n)}-(4m+4){H}_{n-1, {\rm tot}}^{(n-1)}.
\end{equation}
Plugging Eqs.~(\ref{ES12}) and~(\ref{ES13}) into Eq.~(\ref{ES6}) leads to
\begin{eqnarray}\label{ES14}
T_{n, {\rm tot}}^{(n)}&=&\frac{(2m+2)^{n-1}(1+2m)(3+4m)(1-\theta)}{4+4m-2\theta-4m\theta}\nonumber\\
&&+\frac{1+\theta}{4+4m-2\theta-4m\theta}{H}_{n, {\rm tot}}^{(n)}.
\end{eqnarray}
Dividing both sides of Eq.~(\ref{ES14}) by $N_n-1=(2m+2)^n$,  we
arrive at an accurate formula for the ATT $\langle T\rangle_n$, which reads
\begin{eqnarray}\label{ES15}
\langle T\rangle_n &=&\frac{(1+2m)(3+4m)(1-\theta)}{(2m+2)(4+4m-2\theta-4m\theta)}\nonumber\\
&&+\frac{1+\theta}{4+4m-2\theta-4m\theta}\langle H\rangle_n \nonumber\\
&=&\frac{(4m^2+4m+1)(1+\theta)}{2(4m^2+7m+3)(4+4m-2\theta-4m\theta)}2^n(2m+2)^n\nonumber\\
&&+\frac{(16m^2+16m+3)(1+\theta)}{4(4m^2+7m+3)(4+4m-2\theta-4m\theta)}2^n\nonumber\\
&&+\frac{(1+2m)[(8-8\theta)m^2+(11-13\theta)m+(4-5\theta)]}{(4m^2+7m+3)(4+4m-2\theta-4m\theta)}.\nonumber\\
\end{eqnarray}



We continue to express ATT $\langle T \rangle_n$ in terms  of
the network size $N_n$, with an aim to obtain the dependence relation of
$\langle T \rangle_n$ on $N_n$. From $N_n=(2m+2)^{n}+1$, we have
$2^{n}=(N_n-1)^{\ln 2/ \ln (2m+2)}$. Therefore, Eq.~(\ref{ES15}) can
be rewritten as
\begin{small}
\begin{eqnarray}\label{ES16}
& & \langle T \rangle_n \nonumber\\ &=&\frac{(4m^2+4m+1)(1+\theta)}{2(4m^2+7m+3)(4+4m-2\theta-4m\theta)}(N_n-1)^{1+\ln 2/ \ln (2m+2)} \nonumber\\
&&+\frac{(16m^2+16m+3)(1+\theta)}{4(4m^2+7m+3)(4+4m-2\theta-4m\theta)}(N_n-1)^{\ln 2/ \ln (2m+2)}\nonumber\\
&&+\frac{(1+2m)[(8-8\theta)m^2+(11-13\theta)m+(4-5\theta)]}{(4m^2+7m+3)(4+4m-2\theta-4m\theta)},\nonumber\\ \end{eqnarray}
\end{small}
which provides an exact dependence relation of $\langle T \rangle_n$ on $N_n$ and parameter $\theta$. For very large networks, i.e., $N_n\to\infty$, $\langle T\rangle_n$ has the following dominating term:
\begin{small}
\begin{eqnarray}\label{ES17}
& & \langle T \rangle_n \nonumber\\ &\approx&\frac{(4m^2+4m+1)(1+\theta)}{2(4m^2+7m+3)(4+4m-2\theta-4m\theta)}(N_n-1)^{1+\ln 2/ \ln (2m+2)} \nonumber\\
&\sim & \xi(\theta)(N_n)^{1+\ln 2/ \ln (2m+2)} \,,
\end{eqnarray}
\end{small}
where $\xi(\theta)=\frac{(4m^2+4m+1)(1+\theta)}{2(4m^2+7m+3)(4+4m-2\theta-4m\theta)}$.

Form Eq.~(\ref{ES17}) we can observe that in the whole range of $0 \leq \theta \leq 1$, the ATT $\langle T\rangle_n$ scales superlinearly with the system size $N_n$, with the exponent $1+\ln 2/ \ln (2m+2)$ independent of parameter $\theta$. Thus, the inclusion of next-nearest-neighbor jumps, controlled by parameter $\theta$, has a negligible effect on the leading behavior of ATT. However, as shown in Eq.~(\ref{ES17}), the parameter $\theta$ can significantly modify the prefactor $\xi(\theta)$ of the dominatant term for ATT. Concretely, $\xi(\theta)$ is an increasing function of $\theta$. When $\theta$ grows from 0 to 1, the prefactor $\xi(\theta)$ grows from $\frac{4m^{2}+4m+1}{2(4m^{2}+7m+3)(4m+4)}$ to $\frac{4m^{2}+4m+1}{2(4m^{2}+7m+3)}$, implying that the incorporation of next-nearest-neighbor jumps can enhance the transportation efficiency in a significant way. For the two limiting cases of $\theta=0$ and $\theta=1$, the ATT for the former is only $\frac{1}{4m+4}$  of that for the latter.

\section{Full spectrum of fundamental matrix}

In this section, we study the eigenvalues of the fundamental matrix $M_n$ for the trapping problem considered above. We will obtain all eigenvalues as well as their multiplicities. Moreover, we will show that the largest eigenvalue has the same leading scaling as that of  the ATT $\langle T\rangle_n$.
For this purpose, we introduce a matrix $T_n$ defined by $T_n=M_n^{-1}$. Let $\lambda_i(n)$ and $\sigma_i(n)$, where $i=1,2,\ldots,N_n-1$, be the respective eigenvalues of $T_n$ and $M_n$, satisfying $\lambda_1(n)\leq \lambda_2(n)\leq \lambda_3(n) \ldots \leq\lambda_{N_n-1}(n)$ and  $\sigma_1(n)\geq \sigma_2(n) \geq  \sigma_3(n) \geq \ldots \geq\sigma_{N_n-1}(n)$. Then, we have $\lambda_i(n)=1/\sigma_i(n)$. Thus, in order to find the eigenvalues of matrix $M_n$, we only need to determine the eigenvalues for $T_n$.

\subsection{Eigenvalue spectrum for case of $\theta=1$}

We first compute the eigenvalue of $T_n$ for the special case of $\theta=1$. For this case, we use $\Gamma_n$ to denote $T_n$. It is easy to see that for $\theta=1$, the transition probability matrix $P_n$ becomes $P_n=D_n^{-1}A_n$, where $D_n$ is the diagonal degree matrix of $F_n$ with its $i$th diagonal entry being $d_i(n)$. Thus, for $\theta=1$, the $(i,j)$ entry of $\Gamma_n=I-\bar{P}_n$ reduces to the following form:
\begin{equation}\label{T4}
\Gamma_n (i,j)=\begin{cases} 1, & i=j,\\
  -\frac{A_n (i,j)}{d_i (n)}, & i\neq j\,.
\end{cases}
\end{equation}
For the sake of convenience, in the sequel, we use $I$ to denote the identity matrix of approximate order. By definition, the problem of finding eigenvalues of $\Gamma_n$ is equivalent to determine the roots of the characteristic polynomial $\xi_n(\mu)=\det (\mu I-\Gamma_n)$ of $\Gamma_n$. Next, we will derive a recursive relationship for the characteristic polynomial $\xi_n(\mu)$ and $\xi_{n-1}(\mu)$, based on which we will determine all eigenvalues of $\Gamma_n$ from those corresponding to the previous iteration.

By construction, matrix $\Gamma_n$ can be written in a block form:
\begin{equation}\label{T5}
\Gamma_n=\left[\begin{array}{cccc}
\Gamma_{\alpha,\alpha} & \Gamma_{\alpha, \beta} \\
\Gamma_{\beta,\alpha} & \Gamma_{\beta, \beta}
\end{array}
\right]
=\left[\begin{array}{cccc}
I & \Gamma_{\alpha, \beta} \\
\Gamma_{\beta,\alpha} & I
\end{array}
\right],
\end{equation}
where the fact that both $\Gamma_{\alpha,\alpha}$ and $\Gamma_{\beta, \beta}$ are identity matrixes is applied. Then,
\begin{eqnarray}\label{X1}
\xi_{n}(\mu)&=& \left|\begin{array}{cccc}
(\mu-1) I  & -\Gamma_{\alpha, \beta} \\
-\Gamma_{\beta,\alpha} & (\mu-1) I
\end{array}
\right|\nonumber \\
&=& \left|\begin{array}{cccc}
(\mu-1) I- \frac{\Gamma_{\alpha,\beta} \Gamma_{\beta,\alpha}}{\mu-1}& 0 \\
-\Gamma_{\beta,\alpha} &  (\mu-1)I
\end{array}
\right|\nonumber \\
&=& (\mu-1)^{(2m+1)(2m+2)^{n-1}} \left|(\mu-1) I - \frac{\Gamma_{\alpha,\beta} \Gamma_{\beta,\alpha}}{\mu-1}\right|\nonumber \\
&=& (\mu-1)^{2m(2m+2)^{n-1}} \det \left((\mu-1)^2 I- \Gamma_{\alpha,\beta} \Gamma_{\beta,\alpha}\right). \nonumber \\
\end{eqnarray}
In Appendix~\ref{AppA}, we prove that
\begin{equation}\label{T6}
\Gamma_{\alpha,\beta}\Gamma_{\beta,\alpha}=I-\frac{1}{2m+2}\Gamma_{n-1}\,.
\end{equation}
Plugging Eq.~(\ref{T6}) into Eq.~(\ref{X1}) gives
\begin{small}
\begin{eqnarray}\label{X2}
\xi_{n}(\mu)&=& (\mu-1)^{2m(2m+2)^{n-1}} \det \left( (\mu^2 -2\mu) I +\frac{1}{2m+2}\Gamma_{n-1}\right) \nonumber \\
&=& \frac{(\mu-1)^{2m(2m+2)^{n-1}}}{(-2m-2)^{(2m+2)^{n-1}}} \det\left((2m+2)(2\mu-\mu^2)I -\Gamma_{n-1}\right) \nonumber \\
&=& \frac{(\mu-1)^{2m(2m+2)^{n-1}}}{(-2m-2)^{(2m+2)^{n-1}}} \xi_{n-1}\left((2m+2)(2\mu-\mu^2)\right)\,,
\end{eqnarray}\end{small}
which reveals the relationship between $\xi_n(\mu)$ and $\xi_{n-1}(\mu)$, allowing to express the eigenvalues of $\Gamma_n$ in terms of those of $\Gamma_{n-1}$.

Now we show how to obtain the eigenvalues of  $\Gamma_n$ from those of $\Gamma_{n-1}$.  Let $\mu_1(n-1)$, $\mu_2(n-1)$, $\cdots$, $\mu_{N_{n-1}-1}(n-1)$ be the $N_{n-1}-1$ eigenvalues of $\Gamma_{n-1}$. Then, the characteristic polynomial $\xi_{n-1}(\mu)$ of $\Gamma_{n-1}$ can be written as
\begin{equation}\label{T7}
\xi_{n-1}(\mu)=\prod_{i=1}^{N_{n-1}-1}(\mu -\mu_i(n-1))\,,
\end{equation}
substituting which into Eq.~(\ref{X2}) yields
\begin{equation}\label{X3}
\xi_{n}(\mu)=\frac{(\mu-1)^{2m(2m+2)^{n-1}}}{(-2m-2)^{(2m+2)^{n-1}}}\xi_{n-1}(\phi(\mu))\,,
\end{equation}
where
\begin{equation}\label{X3a}
\phi(\mu)=(2m+2)\left(2\mu-\mu^2\right)\,.
\end{equation}

Equation~(\ref{X3}) indicates that $1$ is an eigenvalue of $\Gamma_n$ with multiplicity $2m(2m+2)^{n-1}$ and that all other eigenvalue are determined by $\xi_{n-1}(\phi(\lambda))=0$. For each eigenvalue $\mu_i(n-1)$ of $\Gamma_{n-1}$, from Eq.~(\ref{X3}) we have the
following quadratic equation
\begin{equation}\label{T8}
(2m+2)(2\mu-\mu^2 )-\mu_i(n-1)=0\,.
\end{equation}
Solving this quadratic equation in the variable $\mu$ gives rise to two eigenvalues, $\mu_{i,+}(n)$ and $\mu_{i,-}(n)$, other than $1$ for matrix $\Gamma_n$:
\begin{eqnarray}\label{X4}
\mu_{i,+}(n)=1 +\sqrt{1-\frac{\mu_i(n-1)}{2m+2}}
\end{eqnarray}
and
\begin{eqnarray}\label{X4a}
\mu_{i,-}(n)=1 -\sqrt{1-\frac{\mu_i(n-1)}{2m+2}}\,,
\end{eqnarray}
both of which keep the degeneracy of its parent $\mu_i(n-1)$.

From above analysis, the number of eigenvalues other than 1 is $2(N_{n-1}-1)=2(2m+2)^{n-1}$, and the number of eigenvalue 1 is $2m(2m+2)^{n-1}$. Then, from the eigenvalues of $\Gamma_{n-1}$, we can completely determined all the $2m(2m+2)^n$ eigenvalues of $\Gamma_n$.
For $\Gamma_{1}$, the set of its eigenvalues includes $1$ with multiplicity of $2m-2$, $1+\sqrt{\frac{m}{m+1}}$ and $1-\sqrt{\frac{m}{m+1}}$ with respective multiplicity being $2$. According to the above argument, all eigenvalues and their multiplicities of $\Gamma_{n}$ ($n\geq2$) can be determined in an iterative way: $1$ is always an eigenvalue with multiplicity of $2m(2m+2)^n$, and any other eigenvalue can be obtained by recursively applying Eqs.~(\ref{X4}) and~(\ref{X4a}) with their multiplicity being the same as that of their parent.

\subsection{Eigenvalue spectrum for arbitrary $\theta$}

After obtaining all the eigenvalues of $\Gamma_{n}$ for the particular case of $\theta=1$, we now determine the eigenvalue spectrum of matrix $T_n$ for arbitrary $\theta$ between $0$ and $1$. Let $\zeta_n (\lambda)=\det(\lambda I - T_n)$ denote the characteristic polynomial of matrix $T_n$.
In what follows we will provide a relationship between $\zeta_n (\lambda)$ and $\xi_{n-1}(\mu)$, from which we will show that all the eigenvalues of $T_{n}$ can be completely determined from those of $\Gamma_{n-1}$.

Note that matrix $T_{n}$ can be written in a block form:
\begin{equation}\label{TP1}
T_{n}=\left[\begin{array}{cccc}
T_{\alpha,\alpha} & T_{\alpha, \beta} \\
T_{\beta,\alpha} & T_{\beta, \beta}
\end{array}
\right]
=\left[\begin{array}{cccc}
T_{\alpha,\alpha} & T_{\alpha, \beta} \\
T_{\beta,\alpha} & I
\end{array}
\right].
\end{equation}
Then,
\begin{small}
\begin{eqnarray}\label{X5}
\zeta_{n}(\lambda)&=& \left|\begin{array}{cccc}
\lambda I -T_{\alpha,\alpha} & -T_{\alpha, \beta} \\
-T_{\beta,\alpha} & (\lambda -1)I
\end{array}
\right|\nonumber \\
&=& \left|\begin{array}{cccc}
\lambda I -T_{\alpha,\alpha}- \frac{T_{\alpha,\beta} T_{\beta,\alpha}}{\lambda-1}& 0 \\
-T_{\beta,\alpha} &  (\lambda -1)I
\end{array}
\right|\nonumber \\
&=& (\lambda-1)^{(2m+1)(2m+2)^{n-1}} \left|\lambda I -T_{\alpha,\alpha}- \frac{T_{\alpha,\beta} T_{\beta,\alpha}}{\lambda-1}\right|\nonumber \\
&=& (\lambda-1)^{2m(2m+2)^{n-1}} \times \nonumber \\
&\quad& \det\left(\left(\lambda^2 - \lambda\right)I - (\lambda-1) T_{\alpha,\alpha} - T_{\alpha,\beta} T_{\beta,\alpha} \right).
\end{eqnarray}
\end{small}
The two matrices $T_{\alpha,\alpha}$ and $T_{\alpha,\beta} T_{\beta,\alpha}$ can be, respectively, expressed in terms of matrix $\Gamma_{n-1}$ as (see Appendix~\ref{AppB} for proof)
\begin{equation}\label{TP2}
T_{\alpha,\alpha}=\theta I+(1-\theta) \Gamma_{n-1}
\end{equation}
and
\begin{equation}\label{TP3}
T_{\alpha,\beta}T_{\beta,\alpha}=\theta I-\frac{\theta}{(2m+2)} \Gamma_{n-1}.
\end{equation}

Inserting Eqs.~(\ref{TP2}) and~(\ref{TP3}) into Eq.~(\ref{X5}) leads to
\begin{eqnarray}\label{X6}
\zeta_{n}(\lambda)
&=&\frac{(\lambda-1)^{2m(2m+2)^{n-1}} }{(2m+2)^{(2m+2)^{n-1}}} \times\nonumber\\
&&[(\lambda-1)(1-\theta)(2m+2)-\theta]^{(2m+2)^{n-1}}\times \nonumber\\
&& \xi_{n-1}\left(\frac{(2m+2)(\lambda^2-\lambda-\lambda \theta)}{(\lambda-1)(1-\theta)(2m+2)-\theta}\right),
\end{eqnarray}
which relates $\zeta_{n}(\lambda)$ to $\xi_{n-1}(\eta(\lambda))$, where $\eta(\lambda)=\frac{(2m+2)(\lambda^2-\lambda-\lambda \theta)}{(\lambda-1)(1-\theta)(2m+2)-\theta}$.
Using Eq.~(\ref{T7}), Eq.~(\ref{X6}) can be recast as
\begin{small}
\begin{eqnarray}\label{X7}
\zeta_{n}(\lambda)&=&\frac{(\lambda-1)^{2m(2m+2)^{n-1}} }{(2m+2)^{(2m+2)^{n-1}}} \displaystyle {\prod_{i=1}^{N_{n-1}-1}}\big \{(2m+2)(\lambda^2-\lambda-\lambda \theta)\nonumber\\
&& -\mu_i(n-1)[(\lambda-1)(1-\theta)(2m+2)-\theta]\big \} .
\end{eqnarray}
\end{small}

From Eq.~(\ref{X7}), one can find all the roots  of $\zeta_{n}(\lambda)$, which are the eigenvalues of matrix $T_n$. First, $1$ is a root of $\zeta_{n}(\lambda)$ with multiplicity $2m(2m+2)^{n-1}$. While for other roots different from $1$, they can be derived from the eigenvalues of $\Gamma_{n-1}$. For each eigenvalue $\mu_i(n-1)$ of $\Gamma_{n-1}$, solving the following quadratic equation in variable $\lambda$:
\begin{equation}\label{TP5}
(2m+2)(\lambda^2-\lambda-\lambda \theta)-\mu_i(n-1)[(\lambda-1)(1-\theta)(2m+2)-\theta]= 0 ,
\end{equation}
generates two eigenvalues for $T_n$ unequal to 1, $\lambda_{i,+}(n)$ and $\lambda_{i,-}(n)$,
given separately by
\begin{small}
\begin{widetext}
\begin{eqnarray}\label{TP7}
\lambda_{i,+}(n) &=& \frac{ (1 + m) (1 - \theta) \mu_i(n-1) + (1 + m) (1 + \theta)}{2m+2}+ \nonumber\\
&& \frac{\sqrt{(1 + m)^2 (1-\theta)^2 {[\mu_i(n-1)]^2}  + (1 + m)^2 (1 + \theta)^2 -2 (1 + m) [1 + m (1-\theta)^2 - (1 - \theta) ]\mu_i(n-1)}}{2m+2}\,
\end{eqnarray}
and
\begin{eqnarray}\label{TP8}
\lambda_{i,-}(n) &=& \frac{ (1 + m) (1 - \theta) \mu_i(n-1) + (1 + m) (1 + \theta)}{2m+2}- \nonumber\\
&& \frac{\sqrt{(1 + m)^2 (1-\theta)^2 {[\mu_i(n-1)]^2}  + (1 + m)^2 (1 + \theta)^2 -2 (1 + m) [1 + m (1-\theta)^2 - (1 - \theta) ]\mu_i(n-1)}}{2m+2}\,,
\end{eqnarray}
\end{widetext}
\end{small}
with both $\lambda_{i,+}(n)$ and $\lambda_{i,-}(n)$ having the same multiplicity as that of $\mu_i(n-1)$. In an analogous way, we can verify that all eigenvalues of $T_n$ and their degeneracies can be found by using Eqs.~(\ref{TP7}) and~(\ref{TP8}).

Since there exists a one-to-one relation between the eigenvalues of $T_n$ and the fundamental matrix $M_n$, we thus have also obtained the full eigenvalue spectrum of $M_n$.

\subsection{The largest eigenvalue}

In the above, we have determined all eigenvalues for the inverse $T_n$ of the fundamental matrix $M_n$ and thus all eigenvalues of $M_n$. Here we continue to estimate the largest eigenvalue of the fundamental matrix $M_n$, which is actually equal to the reciprocal of the smallest eigenvalue for matrix $T_n$, denoted by $\lambda_{\rm min}(n)$. We will show that the ATT $\langle T \rangle_n$ has the same leading behavior as that of the reciprocal of $\lambda_{\rm min}(n)$.

We first consider special situation of $\theta=1$, and use $\mu_{\rm min}(n)$ to denote the smallest eigenvalue of matrix $\Gamma_n$. According to the computation process for eigenvalues of $\Gamma_n$, especially Eq.~(\ref{X4a}), the smallest eigenvalue of $\mu_{\rm min}(n)$ satisfies the following recursive relation
\begin{equation}\label{S2}
\mu_{\rm min}(n)=1 -\sqrt{1-\frac{\mu_{\rm min}(n-1)}{2m+2}}.
\end{equation}
Using Taylor's formula, we have
\begin{equation}\label{S3}
\mu_{\rm min}(n)\approx 1-\left(1-\frac{1}{2}\frac{\mu_{\rm min}(n-1)}{2m+2}\right)=\frac{\mu_{\rm min}(n-1)}{4m+4} .
\end{equation}
Considering the initial condition $\mu_{\rm min}(1)=1-\sqrt{\frac{m}{m+1}}$, Eq.~(\ref{S3}) can be solved by induction to yield
\begin{equation}\label{S4}
\mu_{\rm min}(n) \simeq \frac{1-\frac{\sqrt{m^2+m}}{m+1}}{(4m+4)^{n-1}}.
\end{equation}

For arbitrary $\theta$ in the interval $[0,1]$, from Eq.~(\ref{TP8}) it is easy to see that the smallest eigenvalue $\lambda_{\rm min}(n)$ of $T_n$ can obtained from $\mu_{\rm min}(n-1)$ via relation
\begin{small}
\begin{widetext}
\begin{eqnarray}\label{S5}
\lambda_{\rm min}(n) &=& \frac{ (1 + m) (1 - \theta) \mu_{\rm min}(n-1) + (1 + m) (1 + \theta)}{2m+2}- \nonumber\\
&& \frac{\sqrt{(1 + m)^2 (1-\theta)^2 {[\mu_{\rm min}(n-1)]^2}  + (1 + m)^2 (1 + \theta)^2 -2 (1 + m) [1 + m (1-\theta)^2 - (1 - \theta) ]\mu_{\rm min}(n-1)}}{2m+2}.
\end{eqnarray}
\end{widetext}
\end{small}
Again, using Taylor's formula in Eq.~(\ref{S5}), we have
\begin{eqnarray}\label{S6}
\lambda_{\rm min}(n) &\approx&
\frac{2 + 2 m (1 - \theta) - \theta}{(2m+2) (1 + \theta)}\mu_{\rm min}(n-1) \nonumber\\
&\simeq& \frac{[4 + 4 m (1 - \theta) - 2\theta]\left(1-\frac{\sqrt{m^2+m}}{m+1}\right)}{(1 + \theta)(4m+4)^{n-1}}.
\end{eqnarray}
By comparing Eqs.~(\ref{ES15}) and~(\ref{S6}), we can observe that, as expected, the leading behavior for the reciprocal of $\lambda_{\rm min}(n)$ is identical to that of the dominant term for ATT $\langle T \rangle_n$, signaling that the trapping efficiency is characterized by the largest eigenvalue of the associated fundamental matrix $M_n$.

\section{Conclusions\label{Con}}

In this paper, we have presented an analytical study on random walks in a class of scale-free fractal networks, which incorporate both nearest-neighbor and non-nearest-neighbor hopping. We have focused on a particular case of random walks with a single trap placed on the central hub node. By using two different methods, we have deduced two expressions for the MFPT to the trap, which are equivalent to each other. Moreover, we have determined all the eigenvalues and their multiplicities of the fundamental matrix of the random walk, and demonstrated that the largest eigenvalue exhibits the same dominant behavior as that of the MFPT, which validates our computation for the full eigenvalues. The obtained results indicate that the inclusion of non-nearest-neighbor jumps can significantly modify the prefactor of MFPT to the trap. It should be mention that although we only studied a special case that the trap is the central node, the result is similar, when the trap is placed at another node. Our work enables a better understanding of the effect of non-nearest-neighbor hopping on the dynamics of random walks.

\subsection*{Acknowledgment}

The authors thank Bin Wu for his assistance in preparing this manuscript. This work was supported by the National Natural Science Foundation of China under Grants No. 11275049.

\appendix

\section{Derivation of Eq.~(\ref{T6})}\label{AppA}

In order to prove Eq.~(\ref{T6}), it suffices to show that their corresponding entries of two matrices $\Gamma_{\alpha,\beta}\Gamma_{\beta,\alpha}$ and $I-\frac{1}{(2m+2)}\Gamma_{n-1}$ on both sides are equal to each other. For simplicity, let  $Q_n=\Gamma_{\alpha,\beta}\Gamma_{\beta,\alpha}$ and $R_n=I-\frac{1}{(2m+2)}\Gamma_{n-1}$. Obviously, The entries $R_n(i,j)$ of $R_n$ are: $R_n(i,i)=-\frac{1}{2m+2} \Gamma_{n-1}(i,j)$ for $i\neq j$ and $R_n(i,j)=\frac{2m+1}{2m+2}$ for $i= j$. For $Q_n$, its entries $Q_n(i,j)$ can be determined as follows.

If $i=j$, the diagonal entry of $Q_n$ is
\begin{eqnarray}\label{Append01}
Q_n(i,i)&=& \sum_{k\in \beta} \left[\frac{A_{n}(i,k)}{d_i(n)}\cdot\frac{A_{n}(k,i)}{d_k(n)}\right] \nonumber \\
&=& \frac{1}{d_i(n)}\sum_{\substack {A_{n}(i,k)=1 \\ k\in\beta} } \frac{1}{d_k(n)} \nonumber \\
&=& \frac{1}{(m+1) d_i(n-1)}\left[\frac{m d_i(n-1)}{1} + \frac{d_i(n-1)}{2}\right] \nonumber \\
&=& \frac{2m+1}{2m+2}=R_n(i,i)\,,
\end{eqnarray}
where $d_i(n)=(m+1)d_i(n-1)$ has been used.

If $i\neq j$, the non-diagonal entry of $Q_n$ is
\begin{eqnarray}\label{Append02}
 Q_n(i,j)
&=& \sum_{k\in \beta} \left[\frac{A_{n}(i,k)}{d_i(n)}\cdot\frac{A_{n}(k,j)}{d_k(n)}\right]  \nonumber \\
&=& \sum_{\substack{A_{n}(i,k) = 1 \\ A_{n}(k,j) = 1} } \frac{1}{(m+1) d_i(n-1) d_k(n)} \nonumber \\
&=& \frac{ A_{n-1} (i,j)}{(2m+2) d_i(n-1)} \nonumber \\
&=& -\frac{1}{2m+2} \Gamma_{n-1}(i,j)=R_n(i,j).
\end{eqnarray}
Equations~(\ref{Append01}) and~(\ref{Append02}) lead to Eq.~(\ref{T6}).

\section{Derivation of Eqs.~(\ref{TP2}) and~(\ref{TP3})}\label{AppB}

We first prove Eq.~(\ref{TP2}), which provides an expression of $T_{\alpha,\alpha}$ in terms of $\Gamma_{n-1}$. Notice that the diagonal elements of $T_{\alpha,\alpha}$ are all $1$. For a non-diagonal element $T_{\alpha,\alpha}(i,j)$ where $i\neq j$, according to Eqs.~(\ref{T2}) and~(\ref{T4}), we have
\begin{equation}
T_{\alpha,\alpha}(i,j)=\frac{(1-\theta) A_{n-1} (i,j)}{d_i (n-1)}=(1-\theta)\Gamma_{n-1}(i,j).
\end{equation}
Recalling the fact that all the diagonal elements of $\Gamma_{n-1}$ are $1$, it is easy to get
Eq.~(\ref{TP2}).

We proceed to prove Eq.~(\ref{TP3}). To this end, let $\tilde{Q}_n$ and $\tilde{R}_n$ denote, respectively, the two matrices $T_{\alpha,\beta}T_{\beta,\alpha}$ and $\tilde{R}_n=\theta I-\frac{\theta}{(2m+2)} \Gamma_{n-1}$ on both sides of Eq.~(\ref{TP3}). Then, the proof of Eq.~(\ref{TP3}) is reduced to proving the equivalence of
the corresponding entries of $\tilde{Q}_n$ and $\tilde{R}_n$. For matrix $\tilde{R}_n$, it is evident that its diagonal and non-diagonal are $\tilde{R}_n(i,i)=\frac{\theta(2m+1)}{2m+2}$ and $\tilde{R}_n(i,j)=-\frac{\theta}{2m+2} \Gamma_{n-1}(i,j)$, respectively. While for matrix $\tilde{Q}_n$, its entries $\tilde{Q}_n(i,j)$ can be determined in a similar way as those of $Q_n$ for the case of $\theta=1$.

The diagonal entry of $\tilde{Q}_n$ is
\begin{eqnarray}\label{Append03}
\tilde{Q}_n(i,i)&=& \sum_{k\in \beta} \left[\frac{\theta A_{n}(i,k)}{d_i(n)}\cdot\frac{A_{n}(k,i)}{d_k(n)}\right] \nonumber \\
&=& \frac{\theta}{d_i(n)}\sum_{\substack {A_{n}(i,k)=1 \\ k\in\beta} } \frac{1}{d_k(n)} \nonumber \\
&=& \frac{\theta(2m+1)}{2m+2}=\tilde{R}_n(i,i)\,;
\end{eqnarray}
and the non-diagonal element of $\tilde{Q}_n$ is
\begin{eqnarray}\label{Append04}
\tilde{Q}_n(i,j)
&=& \sum_{k\in \beta} \left[\frac{\theta A_{n}(i,k)}{d_i(n)}\cdot\frac{A_{n}(k,j)}{d_k(n)}\right]  \nonumber \\
&=& \sum_{\substack{A_{n}(i,k) = 1 \\ A_{n}(k,j) = 1} } \frac{\theta}{(m+1) d_i(n-1) d_k(n)} \nonumber \\
&=& \frac{\theta A_{n-1} (i,j)}{(2m+2) d_i(n-1)} \nonumber \\
&=& -\frac{\theta}{2m+2} \Gamma_{n-1}(i,j)=\tilde{R}_n(i,j)\,,
\end{eqnarray}
which, together with Eq.~(\ref{Append03}), completes the proof of Eq.~(\ref{TP3}).


%

\end{document}